\documentclass[a4paper,11pt]{article}
\usepackage{aaskaiid}
\usepackage{xspace}
\usepackage{stackengine}
\usepackage{orcidlink}
\def\cgs{\rm{erg}\ \rm{s^{-1}}\ \rm{cm^{-2}}}
\def\ergs{\rm{erg}\ \rm{s^{-1}}}
\def\Msunyr{\rm{M}_\odot\ \rm{yr}^{-1}}

\def\Msun{M$_\odot$}

\newcommand{\NeIV}{{[Ne\,{\sc iv}]\,}}

\newcommand{\NeV}{{[Ne\,{\sc v}]\,}}

\newcommand{\OIIIl}{{[O\,{\sc iii}]\,$\lambda$}}

\newcommand{\CIV}{{C\,{\sc iv}\,}}

\newcommand{\CIVl}{{[C\,{\sc iv}\,$\lambda$}\xspace}
\newcommand{\HeII}{{He\,{\sc ii}\,}}
\newcommand{\HeIIl}{{He\,{\sc ii}\,$\lambda$}}

\newcommand{\Ha}{H$\alpha$\,}

\newcommand{\Hb}{H$\beta$\,}

\def\cgs{\rm{erg}\ \rm{s^{-1}}\ \rm{cm^{-2}}}
\def\ergs{\rm{erg}\ \rm{s^{-1}}}

\title{Unveil the nature of JWST-AGN and Little Red Dots with SKAO continuum surveys}
\ShortTitle{Unveil the nature of JWST AGN and LRD with the SKAO}

\author[1,2,3]{Giovanni Mazzolari\orcidlink{0009-0005-7383-6655}}
\ShortName{Mazzolari et al.} % shortened name list for header 
\author[4]{Dharam V. Lal\orcidlink{0000-0001-5470-305X}}
\author[5]{Isabella Prandoni\orcidlink{0000-0001-9680-7092}}
\author[2]{Roberto Gilli\orcidlink{0000-0001-8121-6177}}
\author[6,7.8]{Roberto Maiolino\orcidlink{0000-0002-4985-3819}}
\author[1]{Hannah Übler\orcidlink{0000-0003-4891-0794}}
\author[2]{Ivan Delvecchio\orcidlink{0000-0001-8706-2252}}
\author[3]{Marcella Brusa\orcidlink{0000-0002-5059-6848}}
\author[2]{Marco Mignoli\orcidlink{0000-0002-9087-2835}}

\affiliation[1]{Max-Planck-Institut für extraterrestrische Physik (MPE), Gießenbachstraße 1, 85748 Garching, Germany}
\affiliation[2]{Istituto Nazionale di Astrofisica (INAF) - Osservatorio di Astrofisica e Scienza dello Spazio (OAS), via Gobetti 101, I-40129
Bologna, Italy}
\affiliation[3]{Dipartimento di Fisica e Astronomia (DIFA), Università di Bologna, via Gobetti 93/2, I-40129 Bologna, Italy}
\emailAdd{gmazzolari@mpe.mpg.de}
\affiliation[4]{National Centre for Radio Astrophysics - Tata Institute of Fundamental Research, Post Box 3, Ganeshkhind P.O., Pune 411007, India}
\emailAdd{dharam@ncra.tifr.res.in}
\affiliation[5]{INAF - Istituto di Radioastronomia, Via Gobetti 101, I-40129 Bologna, Italy}
\affiliation[6]{Kavli Institute for Cosmology, University of Cambridge, Madingley Road, Cambridge, CB3 0HA, UK}
\affiliation[7]{Cavendish Laboratory, University of Cambridge, 19 JJ Thomson Avenue, Cambridge, CB3 0HE, UK}
\affiliation[8]{Department of Physics and Astronomy, University College London, Gower Street, London WC1E 6BT, UK}

\abstract{
The advent of JWST has revealed a large population of Active Galactic Nuclei (AGN) at $z>4$, which are $\sim1$ dex more abundant than previously expected,  including also the enigmatic population of Little Red Dots (LRDs). Remarkably, the vast majority of JWST-discovered AGN and LRDs are not detected in X-rays, and most of them also show faint rest-frame UV continua and faint high-ionization emission lines, as well as unusually faint emission in the Mid and Far infrared. Recent studies investigating their radio properties have reported no significant detections, even in deep stacking analyses, reaching sensitivities of 0.5-0.1 $\mu$Jy at $z\sim 5-6$, corresponding to $L_{R}\lesssim 10^{39}\rm \ erg\ s^{-1}$. While these non-detections may be consistent with a standard radio-quiet nature, some results suggest that the radio emission might instead be significantly suppressed by other physical phenomena. Three main scenarios have been proposed in the literature to explain the physical properties of these objects across the electromagnetic spectrum: Compton-thick absorption by a broad-line region with high covering-factor, intrinsically weak emission driven by high accretion rates, or the presence of a cocoon of dense ionized gas that produces strong scattering effects. The unprecedented sensitivity of SKAO will enable the detection of the radio emission of these AGN in all three cases. Because each scenario is expected to produce distinct radio signatures, future SKAO continuum surveys will be able to distinguish between them, uncovering the physical processes responsible for their peculiar properties. Observations spanning a wide range of integration times (1–1000 hours) and frequencies with SKA-Mid and SKA-Low (0.2–11 GHz) will allow us to characterize these objects from the local Universe to high redshift, investigate possible radio variability, and test alternative scenarios to black hole accretion.}

\begin{document}
\maketitle

\section{Introduction}\label{sec:intro}
Recent studies, exploiting both spectroscopic and photometric JWST data, have revealed a large population of Active Galactic Nuclei (AGN) at $z>4$ \citep{Kocevski25_LRD,ubler23,ubler23b,Matthee23,Maiolino23a,Maiolino23c,Greene24_LRD,Bogdan23,Goulding23,Kokorev23,Furtak23,Juodzbalis24,Scholtz23b,Mazzolari24c,Chisholm24,Juodzbalis25_AGNsample,Taylor2024_AGN}, providing the unique opportunity to study the properties of Supermassive Black Holes (SMBH) and the AGN-galaxy coevolution, since very early times.\\
These works revealed an abundance of active black holes (BHs) in the early universe, contrary to expectations from pre-JWST results.
%\cite{yang23}, taking advantage of the JWST-MIRI photometry of the Cosmic Evolution Early Release Science Survey \citep[CEERS;][]{Finkelstein22}, investigated the AGN population using SED modeling, and found a BHARD at $z>3$ $\sim 0.5$ dex higher than what was expected from previous X-ray AGN studies \citep{vito16,vito18}, and more in line with theoretical models (see Fig.~\ref{fig:JWST_BHRD} to be compared with Fig.\ref{fig:BHARD}). Similarly, \cite{Akins24}, by selecting AGN candidates with NIRCam and MIRI color-color diagnostics on the 0.54 deg$^2$ of the COSMOS-Web survey, derived an almost flat BHARD from $z\sim 2-9$ (even if galaxy contaminants can artificially increase the BHARD). %\cite{Lyu23} performing a multi-wavelength AGN selection also involving the JWST/MIRI data of the SMILES survey \citep{deGraaff24} selected a remarkable fraction of AGN among the MIRI detected sources ($\sim 7\%$) and found a statistically significant increase of the obscured AGN fraction with redshift, confirming the results of other pre-JWST works \citep{Signorini23, gilli22, buchner15, aird15}. \\
Type I AGN have been identified using JWST spectroscopic data looking for sources with broad emission components in the Balmer lines (FWHM$\geq1000$ km s$^{-1}$) and without corrsponding components in the forbidden lines (such as \OIIIl5007), to safely identify sources where the broad components reflect the emission of the rapidly rotating clouds of the broad-line region (BLR) \citep{Maiolino23c, Harikane23, Juodzbalis25_AGNsample, Taylor2024_AGN, Hviding25}. These selections enabled the derivation of Type I AGN luminosity functions at $z>4$ up to 1 dex larger compared to the extrapolations of the X-ray selected AGN \citep[][]{giallongo19} or UV selected AGN \citep{matsuoka19,Banados16,Shen20} luminosity functions.\\
These high-z AGN discovered by JWST are characterized by bolometric luminosities of 1-3 dex fainter compared to the previously known AGN population at the same redshift  ($43<\log (L_{bol}/ \rm erg\ s^{-1})<46$) and are generally powered by BHs typically of $M_{BH}\sim 10^{6-7}$ M$_{\odot}$, placing them among the least-massive BHs known in the early Universe. These faint AGN are probably more representative of the global BH population at high redshifts and, given the importance of retrieving a complete census of the AGN populations at different epochs, they can be crucial in constraining models of BH seeding \citep{Pacucci22,Li23}, to investigate the contribution of AGN to hydrogen reionization \citep{Dayal20,Yung21}, and the early coevolution of galaxies and BHs \citep{Habouzit22, Inayoshi22, Pacucci23,Jones25_MBHMstar}. 

Different studies have revealed that this population of early BHs detected with JWST is generally overmassive relative to the host galaxy stellar mass when compared with the local AGN distribution \citep{ubler23,Maiolino23c, Bogdan23, Furtak23, Juodzbalis24, Juodzbalis25_AGNsample, Jones25_MBHMstar, Ma26_undermasivehost}, and considering the local scaling relations \citep{Reines15, Kormendy13}. This might suggest that the early stages of the BH-host galaxy coevolution can be dominated by a first phase of BH growth, followed by an intense phase of star formation, allowing the host galaxy to catch up with the BH growth and eventually align with the local scaling relations. This could also provide indications on the seeding mechanisms of these BH, potentially preferring heavy-seed scenarios, together with episodes of super-Eddington accretion \citep{Chon26_overmassiveBH,Zhang25_primordialBH, Scholtz_2023_GN-z11,Lupi24, Maiolino25_QSO1a}. However, large uncertainties remain in the precise estimate of the BH-to-stellar-mass ratios of these objects, given the uncertainties of locally calibrated scaling relations used to estimate the BH masses \citep{Reines15, Greene05} and the difficulty of providing reliable stellar masses using SED-fitting decomposition \citep{Choe25_uncertainMstar, Ronayne25_LRDsedfit, Ma26_undermasivehost, Juodzbalis25_AGNsample}.

Among the AGN selected using JWST data a peculiar population have been identified, the so called Little Red Dots (LRD). These sources have compact sizes (unresolved at the JWST PSF, i.e. $<0.1^{\prime\prime}$ ) and are characterized by a "V-shape" spectral energy distribution (SED) with a steep red continuum in the rest-frame optical, blue colors in the rest-frame UV \citep{Kocevski25_LRD,Harikane23,Matthee23,Greene24_LRD,Killi23} and a turnover in correspondence of the Balmer break \citep{Setton25_Vshape, DeGraaff25_BB}. This population was originally selected photometrically from NIRCam images, exploiting its peculiar SED shape and the characteristic compactness. %These photometric selections turned out to contain both AGN, dust-obscured starburst galaxies, and a minor population of brown dwarf \citep{PerezGonzalez23, Akins25_LRD, Labbe25_LRDsample}. 
Remarkably, follow-up spectroscopy of these sources showed that over 70-80\% of photometrically selected LRDs exhibit broad-line Balmer emission, strongly supporting their AGN nature \citep{Greene24_LRD,Kocevski2024_LRD,Hviding25}, and with only a minor contamination from dust-obscured starburst galaxies, and brown dwarf \citep{PerezGonzalez23, Akins25_LRD, Labbe25_LRDsample}. The evidence of the presence of an AGN in these compact and peculiar sources suggested that the compact nuclear accretion must have an impact on their peculiar SED emission and morphologies, with some works suggesting they might represent the first phase of BH activity after seeding \citep{Inayoshi25_LRDfirstBH, Pacucci25_directcollapseLRD,Cenci25_directcollapseLRD}.  

A remarkable feature of high-$z$ AGN discovered by JWST and LRD is that a large fraction of them is undetected in the available deep X-ray images.
Specifically, most of the newly selected AGNs, including Type I AGN, lack any X-ray emission \citep{Maiolino24_X, Yue24, Ananna24, Mazzolari24c}, even if located in fields covered by some of the deepest extragalactic X-ray observations ever performed, such as the Chandra Deep Field South \citep[CDFS][]{luo17} or Chandra Deep Field North \citep[CDFN][]{xue16}, (having X-ray flux limits $f_{2-10 keV}\sim 10^{-18}-10^{-17}\cgs$). The X-ray non-detections persist even after X-ray stacking. Specifically, \cite{Maiolino24_X} derived X-ray luminosity upper limits on the single sources of $\log (L_{2-10~keV}/\rm erg\ s^{-1})\lesssim42-43$ and from the stack of $\log (L_{2-10~keV}/\rm erg\ s^{-1})\lesssim41.5$. By comparing the observed X-ray luminosity upper limits with the expected X-ray luminosities \citep[obtained from the bolometric luminosities using][]{duras20} it was shown that these sources are characterized by a severe X-ray weakness, up to 2-3 dex \citep{Maiolino24_X, Yue24,Ananna2024}. Given the very few exceptions of X-ray AGN detections in the early Universe \citep{Goulding23,Kovacs24,Maiolino24_X}, it is possible that the X-ray weakness could be due to intrinsic properties of high-$z$ AGN \citep{Tortosa26_Xweak, Yue24, Lambrides24}, as it was also suggested for some low-$z$ AGN \citep{simmonds18,Zhang23}. In this view, these AGN might be characterized by a different accretion-disk/coronal structure that can determine, for example, a larger ratio of the optical to X-ray emission ($\alpha_{OX}$) due to a much lower efficiency of the corona in producing X-ray photons \citep{Proga05}, or simply a lack of the corona itself. Another possibility is that these sources are undergoing (or have undergone) a phase of rapid, super-Eddington accretion, collimating the X-ray emission and altering the X-ray corona structure \citep{Pacucci24, Lambrides24, Madau2024_SE, King25, Tortosa26_Xweak}.\\ 
This scenario was also proposed based on their rest-frame UV emission. Indeed, most of the JWST-selected LRDs and Type I AGN do not show the typical high-ionization emission lines observed in the standard Type I AGN, such as \NeV, \NeIV, \CIV, or \HeII \citep{Lambrides24, Deugenio25_irony, Torralba25}. Specifically, \cite{Lambrides24} showed that  \CIVl1548, and \HeIIl1640 are usually undetected in JWST AGN spectra down to $\log (L_{line}/\rm erg\ s^{-1})\lesssim41$, while from standard AGN disk models these lines should have luminosities $\log (L_{line}/\rm erg\ s^{-1})\sim42$. Instead, super-Eddington accretion can explain this weakness, predicting line luminosities that are 2-4 dex lower than standard disk models.\\
Other works, analyzing the lack of X-ray emission in a large sample of high-$z$ sources unambiguously identified as AGN, suggested that their X-ray weakness could also be ascribed to obscuration and to the presence, close to the broad-line region, of an almost spherical distribution of clouds with Compton-thick column densities and very low dust content \citep{Maiolino24_X, Juodzbalis24_rosetta, Juodzbalis25_AGNsample}. The same effect can also be produced by the broad line region clouds themselfs, if distributed with a larger convering factor compared to local, higher luminosities AGN. A low amount of dust in these objects is needed to justify the presence of high S/N broad emission lines (that otherwise would be significantly suppressed) and is supported by the non-detections of the warm dust continuum of these objects at mm wavelengths \citep{Casey24, Casey25, Setton25}, and also by the relatively flat mid-infrared slopes \citep{Setton25, Wang24, Williams24_flatMIR, Delvecchio25}.
In this hypothesis, the observed lack of X-ray emission would be due to the inner Compton-thick gas distributed with large covering factors, to justify the widespread X-ray non-detections of these objects. Single broad line region clouds are already expected to have column densities $N_{H}>10^{23}-10^{24}\, \rm cm^{-2}$ \citep{Zhang17_BLRclouds, Panda20_BLRcloud}, and they are also naturally dust poor, given that they reside within the dust sublimation radius \citep{Risaliti11}.
Furthermore, the presence of a dense distribution of neutral gas was also proposed to justify the detection in 20-30\% of JWST AGN of narrow absorption features in the Balmer or HeI emission lines \citep{Matthee23, Juodzbalis24_rosetta, Juodzbalis25_AGNsample, DEugenio25a, Wang24, Loiacono25}, which needs very high neutral Hydrogen densities (i.e. $n_H\sim10^9-10^{10}$ cm$^{-3}$) to be produced. In this scenario, \cite{Inayoshi2024} demonstrated that such high densities of neutral Hydrogen can also naturally explain the V-shape of LRDs, through absorption in all Balmer lines till $n=\infty$. The explanation of the V-shaped SED of LRD and of their Balmer break in terms of a gas-enshrouded AGN further alleviates the tensions that would arise from interpreting the Balmer break as due to the emission of old stellar populations. In that case, given the compact size of these sources, the large stellar masses would result in enormous stellar densities ($\Sigma_*=10^{11}-10^{12}$\Msun $\rm kpc^{-2}$), even larger than the densest stellar cluster known \citep{Akins25_LRD}, and their number density would further create tensions with the expectations from the $\Lambda$CDM model \citep{Labbe23_lambdaCDM}.

Recently, different works hypothesized that, at the high gas densities expected in this 'gas-embedded' AGN model, also scattering processes may play a crucial role \citep{Rusakov25,Chang25, Naidu25,DeGraaff25_BB, Sneppen26_cocoon}. \cite{Rusakov25} first noted that the broad Balmer line profiles of some of these AGN have an exponential shape (rather than Gaussian), apparently pointing to the effects of electron scattering in dense, but at least partially ionized gas. In this dense gas picture, the gas is optically thick to the Balmer transitions, and the \Ha emission may be partially produced in outer layers of the envelope \citep{Begelman26_quasistar}, while scattering might happen in a much inner and thinner region where $n_{e}\sim 10^9 \rm cm^{-3}$ \citep{Sneppen26_cocoon}. \cite{Chang25} suggested that a mix of resonant and electron scattering can explain, at least in part, the emission line shapes and absorption features observed in these objects. While \cite{Scholtz26_expo} showed that the exponential profiles of Balmer lines are not a sufficient condition to claim the presence of electron scattering in these sources, there are also other elements pointing towards a role played by scattering (at least in some LRDs), such as the high Balmer decrements (\Ha/\Hb$\gtrsim10$), hard to explain in terms of obscuration \citep{DeGraaff25_BB}. 

As presented above, the population of AGN discovered by JWST appears to exhibit peculiar physical properties that are uncommon among AGN discovered before JWST. Radio observations may provide important clues about the processes underlying the observed properties of this new population of BHs and help disentangle the main scenarios proposed to explain them: dense neutral gas obscuration, intrinsically different accretion properties, and scattering processes related to ionized dense gas. AGN radio emission is generally much less affected by obscuration than the X-ray emission \citep{Mazzolari24a, Ricci24} and a high level of radio emission can be produced by processes associated with SMBH activity \citep[e.g. coronal activity, shocks, small-scale radio jets, see][for a review]{panessa19} even in systems that are not classical radio loud (RL) AGN, as demonstrated by the effectiveness of 'radio-excess' AGN selection techniques in selecting also radio-quiet (RQ) objects \citep{bonzini13,smolcic17b, delvecchio17, whittam22, Mazzolari26_radioCTK}. Therefore, in the obscuration scenario, one might expect to detect these AGNs at radio wavelengths if the observations are sufficiently deep. On the contrary, if the properties described above are due to intrinsically different accretion properties or to super-Eddington accretion (producing an intrinsically different, and possibly weaker, SED), then one might expect to find also a weaker-than-expected emission in the radio band compared to what is suggested by the standard Type I AGN scaling relations \citep[most of highly accreting low redshift BH are indeed found to be radio quiet][]{Kellermann1989,Greene06}. Additionally, large column densities of free electrons may instead produce free-free absorption, which can absorb part of the radio emission (especially at the lower frequencies) and produce distinct radio spectral shapes.
Therefore, in addition to contributing to the overall understanding of the multi-wavelength emission of these sources, radio observations can shed light on the true nature of the physical properties that characterize JWST-discovered AGN.

\section{Failed attempts to detect the radio emission of JWST-discovered AGN} \label{sec:failure}

\begin{figure}[htb]
    \centering
    \includegraphics[width=1\linewidth]{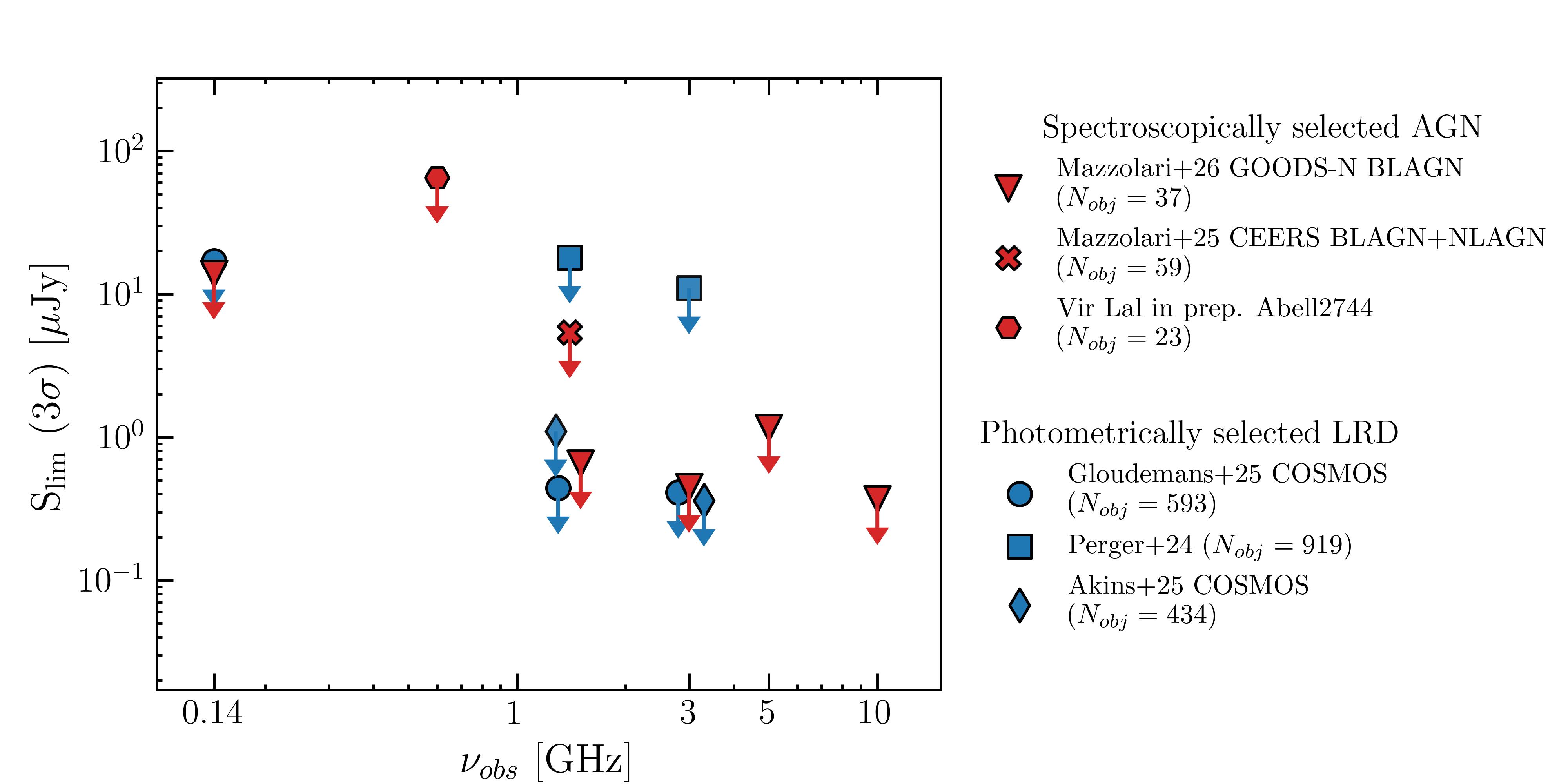}
    \caption{Visual representation of the series of failed attempts to detect the radio continuum emission in AGN discovered by JWST via stacking experiments. Blue points represent the results from stacking experiments of photometrically selected LRDs, while red markers indicate samples of spectroscopically confirmed Type I AGN.  The average redshift of the samples are $5<z<6$.}
    \label{fig:literature}
\end{figure}

In the past year, different works investigated the radio properties of JWST-selected AGN, considering both photometrically selected LRDs and spectroscopically selected Type I and Type II AGN \citep{Perger24, Mazzolari24c, Mazzolari24d, Akins25_LRD, Gloudemans25}. The analyses conducted so far have not succeeded in detecting these sources in any radio band, not even with stacking analyses on the deepest radio observations to date. We summarize the results obtained so far in Table~\ref{tab:samples}. In Fig.~\ref{fig:literature}, we show a compilation of 3$\sigma$ upper limits collected from the literature. 

\begin{table}[h!]
\centering
\begin{tabular}{l l c c c}
\hline
\textbf{Work} & \textbf{Sample} & \textbf{\# of Objects} & \textbf{Frequency [GHz]} & \textbf{flux limit ($3\sigma$) [$\mu$Jy]} \\
\hline
 \cite{Akins25_LRD} & P & 434 & 3  &  0.36\\
 \cite{Akins25_LRD} & P & 434 & 1.28  &  1.08\\
 \cite{Perger24} & P & 919 & 1.4 &  18\\
 \cite{Perger24} & P & 919 & 3 &  11\\
 \cite{Gloudemans25} & P & 593 & 3 &  0.41\\
 \cite{Gloudemans25} & P & 593 & 1.3 &  0.44\\
 \cite{Gloudemans25} & P & 593 & 0.144 &  17\\
 Vir Lal, in prep. & S & 23 & 0.6 &  65\\
 \cite{Mazzolari25_CEERS} & S & 59 & 1.4 &  5.4\\
 \cite{Mazzolari26_radioJWST} & S & 37 & 0.144 &  14\\
 \cite{Mazzolari26_radioJWST} & S & 37 & 1.4 &  0.65\\
 \cite{Mazzolari26_radioJWST} & S & 37 & 3 &  0.45\\
 \cite{Mazzolari26_radioJWST} & S & 37 & 5 &  1.16\\
 \cite{Mazzolari26_radioJWST} & S & 37 & 10 &  0.37\\

\hline
\end{tabular}

\caption{Summary table of the stacking attempts performed so far to detect JWST-discovered AGN and LRDs in radio images. The first column refers to the literature work in which the stacking was performed, the second is 'S' if the sample was spectroscopic and 'P' if photometric, the third refers to the number of objects in the stack, the fourth to the frequency of the radio image, the fifth the the $3\sigma$ upper limit obtained from the stack. }
\label{tab:samples}
\end{table}

\cite{Akins25_LRD} obtained among the deepest constraints, performing a radio stacking analysis of 434 photometrically selected LRDs in the COSMOS-Web field using the radio imaging from MeerKAT at 1.4 GHz \citep{Heywood22} and VLA at 3 GHz \citep{smolcic17a}. From the radio non detections \cite{Akins25_LRD} derived an upper limit to the radio-loudness parameter of these sources \citep[$R_{4400\AA}=L_{5GHz}/L_{4400\AA}$][]{Kellermann1989} of $R_{4400\AA}\leq 10$. Instead, \cite{Perger24} stacked the VLASS and FIRST cutouts of the largest compilation of LRDs, including 919 photometrically selected sources across different fields exploited by JWST. These studies considered the largest sample of sources; however, as mentioned before, none of these were detected as single sources nor via stacking.

Since the photometrically selected LRDs may be contaminated by non-AGN galaxies \citep[see brown dwarfs and red galaxies contamination rates reported in][]{Greene24_LRD,Kocevski2024_LRD,Langeroodi23_BD-LRD}, their  average radio emission might be fainter due to contamination. Therefore, focusing on spectroscopically selected AGN samples might yield more precise and realistic conclusions about the average properties of these sources. In particular, to understand the accretion nature of these sources and the origin of their peculiar emission across the whole SED, it is necessary to verify whether the observed radio upper limits are consistent with the scatter of the expected Type I AGN radio emission \citep[see RQ and RL Type I AGN template in][]{Shang11}. \cite{Mazzolari25_CEERS} investigated the radio counterparts of spectroscopically selected Type II and Type I AGN in the EGS field, but the limited number of sources and the $\sim 10 \mu$Jy depth of the existing radio image \citep{Ivison07} did not allow deriving constraining results on their radio properties from the radio non-detection. \cite{Gloudemans25} looked for radio counterparts of multiple samples of spectroscopically and photometrically selected JWST AGN on COSMOS, GOODS-N, and GOODS-S, and they concluded that their radio non-detections were consistent with a standard radio-quiet AGN nature. Interestingly, they also found one of these AGN to be radio-detected in the COSMOS radio images (rest frame $L_{\rm 3~GHz}=9\times10^{40} \ergs$). However, even if it was selected using LRD selection criteria, this source showed all the typical and classic AGN features: it was X-ray bright ($L_{\rm 2-10~keV}\sim 5\times 10^{44}\ergs$) and its luminosity was consistent with the standard AGN scaling relations. Therefore, also the radio detection of this source was expected, given that it does not share all the other peculiar physical properties of the newly JWST-selected AGN.

Finally, \cite{Mazzolari26_radioCTK} analyzed a sample of 37 Type I AGN in five deep radio images in the GOODS-N field, spanning a frequency range from 144 MHz to 10 GHz. By comparing the observed radio luminosity upper limits of each source with the expected radio luminosities returned by four different correlations between $L_{R}$ and other observed quantities, they found that the single sources' upper limits were still within the scatter of the correlations for RQ sources but were inconsistent with a RL nature of the sources. Instead, considering the results from the stack, they noticed that the observed radio luminosity upper limit corresponding to the stack was significantly lower than expected, suggesting, on average, a weaker-than-expected radio output. Additionally, by computing the probability of these sources being standard RQ AGN but still undetected in all of the investigated radio images, they found probabilities of the order of $10^{-4}$. The observational results, as well as the outcome of this statistical test, suggested that these sources might actually populate the faintest end of the AGN radio-loudness distribution and therefore appear as "radio weak".

One might argue that the non detection of these sources in the radio band could be due to a redshift effect. Indeed, the negative k-correction of the AGN radio spectrum naturally makes the radio detection of AGN at high-z much more complex. The AGN considered in the above-mentioned studies are at $z\sim4-6$, but the same undetections were also retrieved for $z\sim 2-3$ analogs of JWST-discovered AGN \citep{Juodzbalis24_rosetta,Loiacono25, Wang24}, supporting the radio non detection of this kind of sources as an intrinsic property of this population rather than a cosmological effect.

Recently \cite{Rodriguez26_radioLRD} reported the only radio detection of a local LRD originally identified by \cite{Lin26_localLRD} at $z = 0.168$ among the SDSS AGN spectra (J1047+0739). The radio emission is optically thin, based on the in-band spectrum ($\alpha=-0.85$). The rest frame radio luminosity is $L_{\rm 6~GHz}=5\times10^{38} \ergs$, or integrated between $1<\nu_{rest}/\rm GHz<10$ $L_{R}=1.5\times10^{39} \ergs$. These luminosities are a factor of a few lower than those probed by the stacks reported in Table~\ref{tab:samples} and demonstrate that the sensitivities of the next generation of radio surveys are needed to potentially detect higher-z LRDs. The source is detected in two different epochs (2010 and 2018), but the lack of significant time variability ($>20\%$) between the two discard the possibility of radio emission from a signle supernova event and rather suggests that it can be attributed to a small black hole of similar nature of those found in Seyfert galaxies \citep{Rush96_seyfertradio} or by the emission produced by past generations of supernovae. 

%If the radio emission from cosmological LRDs is similar in luminosity to that produced by LLRDs it will be possible to detect them with long integrations with the next generation of centimeter interferometers.
In this Chapter, we want to investigate the detectability of JWST-discovered AGN and LRD with the future surveys that will be performed by the SKAO. Additionally, we want to explore how and if radio emission can help in disentangling between the different scenarios proposed to explain the overall physical properties of this newly discovered class of AGN.

%\subsection{Results using Giant Metrewave Radio Telescope}
%Figure~\ref{fig:A2744-lrds} shows high resolution high sensitivity image of the Abell 2744 cluster.  Abell 2744 is an X-ray luminous, rich merging and hot cluster, with 1.4$-$2.5 $\times$ 10$^{15} M_\odot$ within 1 Mpc \citep{Boschin}. 
%It hosts a central radio halo and a peripheral relic \citep{Orru2007}, which are also seen in our low-frequency radio image at 606 MHz.

%Twenty three LRDs listed in \cite[][see Figure~\ref{fig:A2744-lrds}]{Ananna2024} are background sources, whereas the Abell 2744 cluster is the foreground cluster.  Thus due to high radio emission in the foreground makes us difficult to detect and image LRDs with extremely weak radio emission. 

%\begin{figure}
%    \centering
%    \includegraphics[width=1.0\linewidth]{figures/fig-A2744-lrds.pdf}
%    \caption{High resolution high sensitivity image of 23 LRDs located in the Abell 2744 cluster. The noise level is 0.1 mJy~beam$^{01}$ and the angular resolution is 6.7$^{\prime\prime}$ $\times$ 4.4$^{\prime\prime}$ with a PA = 6.3 $^\circ$. LRD positions are marked by the ``$\circ$'' sign.}
%    \label{fig:A2744-lrds}
%\end{figure}

\section{Advances with SKAO} \label{sec:SKA}
Current attempts to detect the radio emission of JWST-selected AGN at high and low redshifts have been largely unsuccessful, primarily due to the limited sensitivity of existing radio surveys. While some results suggest that, on average, the radio emission from these sources might be unexpectedly faint, current radio surveys and observatories do not permit deeper investigations. The SKAO is planned to conduct the deepest radio continuum observations ever, thereby critically contributing to the detection of radio emission from these sources and enhancing our understanding of their accretion properties.

\subsection{Radio identification and physical characterization} \label{sec:radio_detection}
The first scientific goal we expect to achieve with SKAO observations is to detect the radio emission of these peculiar AGN.
In the following we will discuss the detectability of these sources considering the three scenarios outlined above:
\begin{itemize}
    \item Standard AGN with Compton thick obscuration from neutral gas
    \item Intrinsic weakness of the X-ray and radio emission
    \item Electron scattering producing free-free absorption of radio emission
\end{itemize}

For the three scenarios, we derived the expected radio fluxes as follows. 
%In Fig.~\ref{fig:detectability}, we simulate the expected AGN and host galaxy radio spectrum and compare them with the 5$\sigma$ detection limits that SKA is expected to achieve with 1, 20, and 100 hours of observations.
\subsubsection{Standard, Compton-thick AGN}
For the first scenario, we followed the same approach as in \cite{Latif24}, \cite{Gloudemans25}, and \cite{Mazzolari26_radioJWST}, i.e. using the AGN fundamental plane scaling relation. This class of scaling relations correlates the AGN X-ray luminosity and BH mass with its radio emission, in the form: $\log L_R=\xi_{RX}\,\log L_X + \xi_{RM} \log M_{BH}$. They have been investigated since the beginning of the century \citep{Merloni02,merloni03} and different works showed that such a correlation is valid across different orders of magnitude of the quantities involved (even if the coefficient $\xi_{RX},\, \xi_{RM}$ can vary), suggesting that it could describe a fundamental property of BH accretion \citep{Li08, Dong14, dong21, Wang24_FP, Bariuan22}. In particular, we chose the fundamental plane relation derived for RQ AGN by \cite{Wang24_FP}:
\begin{equation}
\log L_{\mathrm{5~GHz}} = 0.47\,\log L_{\mathrm{X}} + 0.29\,\log M_{\mathrm{BH}} + 17.06.
\label{eq:fund_plane}
\end{equation}

We chose this relation because it was derived considering RQ AGN samples up to $z>4$ and selected from some of the deepest fields in terms of radio and X-ray flux sensitivities. Furthermore, this relation was derived without pre-selecting sources with a specific kind of accretion physics, thereby preventing us from making any a priori assumptions about the accretion mechanism of the JWST-selected AGN. While the fundamental plane relation might not necessarily hold for this specific type of source  (on which it was not possible to test these correlations due to the lack of radio and X-ray detections), with this attempt, we want to provide a reasonable broad range of radio fluxes at which we expect to detect the radio emission of these sources. The RL AGN hypothesis had already been discarded by most of the works in the literature, and therefore, we conservatively assumed only RQ emission. In any case, the presence of strong radio jets in some of these sources would only increase their radio flux. We used Eq.~\ref{eq:fund_plane} to estimate the AGN rest frame 5~GHz luminosity, then we derived the observed radio flux by inverting the equation:
\begin{equation}\label{eq:Lrad}
    \rm L_{5~ GHz}=\frac{4\pi d_L^2 S_{5~GHz}}{(1+z)^{1+\alpha}} \ \ W\, Hz^{-1},
\end{equation} 
where $d_L$ is the luminosity distance, $\alpha$ the radio spectral index, and $S$ is the radio flux.  We assume a flat $\Lambda$CDM universe with $H_{0}=70\ \rm{km s^{-1} Mpc^{-1}}$, $\Omega_{m}=0.3, \Omega_{\Lambda}=0.7$. We assume radio spectra following a powerlaw of the form $S_{\nu}\propto \nu^{\alpha}$, with $\alpha=-0.5$ consistent with the values found in the literature for AGN-dominated radio emission \citep{Wang24_FP, Bariuan22}. Finally, using the same spectral index we derived the radio spectrum at the different frequencies.

We computed the X-ray luminosities needed for Eq.~\ref{eq:fund_plane} in the following way. 
%We considered two different scenarios for the expected AGN radio emission: standard AGN, or intrinsically X-ray weak AGN, for example, as a consequence of a different kind of accretion mechanism, as discussed in Sec.~\ref{sec:intro}. Indeed, we recall that the origin of the weak X-ray and UV emission of these sources is still debated, and it is not yet clear whether it is due to gas obscuration or to an intrinsically different accretion structure. 
We started by considering the median redshift and bolometric luminosity of the samples of JWST-discovered AGN selected so far, i.e. $L_{bol}\sim10^{45}$ erg s$^{-1}$ and $z=5.5$. In the first scenario, the overall AGN emission is expected to be "standard" (i.e., following the typical scaling relations of other Type I AGN), but obscured by Compton-thick hydrogen column densities. Therefore, we derived the expected X-ray luminosity from the bolometric one using the \cite{duras20} relation: 
\begin{equation}
K_X \equiv \frac{L_{\mathrm{bol}}}{L_{2-10 keV}} =
a \left[
1 + \left(
\frac{\log\left(L_{\mathrm{bol}} / L_\odot\right)}{b}
\right)^{c}
\right],
\end{equation}
where we took the a, b, c coefficients reported in Table~1 of \cite{duras20} for the Type I AGN population. For the BH masses, we considered the range $7<\log (M_{BH}/M_{\odot})<8$. 
In the first scenario, while the X-ray emission might be significantly absorbed by the high gas column densities, the radio emission is not expected to be affected by obscuration effects, as it is largely optically thin under these conditions.

\subsubsection{Intrinsic X-ray and radio weakness}
The second scenario invokes an intrinsically weak X-ray emission, originating as a consequence of a different kind of accretion mechanism (like super-Eddington accretion), as discussed in Sect.~\ref {sec:intro}. In this case, we considered an X-ray luminosity $\sim2$ dex fainter than that predicted by the \cite{duras20} relation reported above. This intrinsic X-ray weakness is therefore expected to determine also an intrinsic radio weakness, following Eq.~\ref{eq:fund_plane}. The correlation between X-ray and radio weakness was analyzed in \cite{Mazzolari26_radioJWST}, which discussed the radio non-detections of JWST-discovered AGN in light of results on the origin of X-ray weakness in super-Eddington sources reported in \cite{Pacucci24}.  Critically, analyzing the fundamental plane relation for a sample of 69 highly accreting AGN, \cite{Paul26_fundplaneEdd} showed that, once correcting $L_X$ for X-ray weakness, the $\xi_{RX}$ and $\xi_{RM}$ coefficients describing the fundamental plane relation of these sources are consistent with those of \cite{Wang24_FP} assumed here \citep[see Fig.~5 in][]{Paul26_fundplaneEdd}. For the BH masses, we assumed the same range as in the first scenario.

\subsubsection{Free-free absorption}
Finally, the third scenario invoked to explain the properties of JWST-discovered AGN and LRDs, considers the presence of an ionized hydrogen layer, more internal (and probably also thinner) than the neutral hydrogen one \citep{DeGraaff25_BB}, and responsible for electron and resonant scattering \citep{Chang25}. In this ionized hydrogen layer the scattering processes may give origin the exponential wings observed in the broad Balmer lines of some of these objects, but the presence of large reservoirs of free electrons may also significantly affect the transmission of the radio emission from the central BH. Indeed, radio emission is generally largely unaffected by obscuration and can be observed even from the most obscured environments, contrary to X-rays. However, when large densities of free electrons are available, also radio emission, especially at low frequency, can be absorbed as a consequence of free-free absorption \citep{Odea98, Laor08, Baskin21}. In this scenario, we assumed the radio emission produced by a standard AGN (as in the first scenario), but introducing the effect of free-free absorption in the form: $L_{R}=L_{R}\, e^{-\tau_{ff}}$, where the free-free optical depth $\tau_{ff}$ is defined by:
%%%%%%%%%%%%%%%%%%%%%%%%%%%
\begin{equation}
    \rm \tau_{ff}=3.28\  10^{-7}\times \biggl(\frac{T_e}{10^4 K}\biggr)^{-1.35}
    \biggl(\frac{\nu_{rest}}{GHz}\biggr)^{-2.1}
    \biggl(\frac{EM}{pc\ cm^{-6}}\biggr),
\end{equation}
%%%%%%%%%%%%%%%%%%%%%%%%%%
%%%%%%%%%%%%%%%%%%%%%%%%%%
where $T_e$ is the electron temperature, $\nu_{rest}$ is the rest frame frequency of the radio emission (i.e., $\nu_{rest}=(1+z)\nu_{obs}$), and $EM$ is the emission measure, which is defined as the integral of the square of the free electrons density $n_e^2$ along the line of sight. The $EM$ can be easily expressed in terms of the column density of ionized hydrogen, in the form of $N_e^2~r^{-1}$ where $r$ the thickness of the ionized medium along the line of sight, therefore: 
%%%%%%%%%%%%%%%%%%%%%%%%%%%%
\begin{equation}\label{eq:tau_ff_Ne}
   \rm  \tau_{ff}=3 \times   \ \biggl(\frac{N_e}{10^{22} cm^{-2}}\biggr)^2
    \biggl(\frac{T_e}{10^4 K}\biggr)^{-1.35}
    \biggl(\frac{\nu_{rest}}{GHz}\biggr)^{-2.1}
    \biggl(\frac{r}{1 pc}\biggr)^{-1}
\end{equation}
%%%%%%%%%%%%%%%%%%%%%%%%%%%
in the source rest frame. In particular, in the following we use as a reference: $T_e=5\times10^4$K, $N_e=10^{23}\rm \,cm^{-2}$ and $r=0.01~$pc.\\
It is worth noting that if the broad emission components observed in the Balmer lines of these objects really arise from scattering processes, then the broad lines no longer trace the motion of the gas clouds in the BLR, and therefore the BH masses estimated using the local virial relations can be significantly overestimated, even more than a factor of $\sim$10 \citep{Rusakov25, Chang25, Naidu25}. This is potentially another effect that can contribute to lowering the expected radio emission of JWST-discovered AGN and LRD in this scenario.
However, considering the fundamental plane relation reported in Eq.~\ref{eq:fund_plane}, if the real BH masses are lower by a factor of ten, then the expected radio emission becomes a factor of $\sim2$ lower, given the mild dependence of $L_{rad}$ on $M_{BH}$ ($\log L_{rad}\propto 0.28\times \log M_{BH}$). Given this mild dependence (compared to the huge impact of free-free absorption), we decide to assume the same range of $M_{BH}$ as in the other two scenarios.\\

\begin{figure}
\vspace{-1.02cm}
    \centering
    \includegraphics[width=0.9\linewidth]{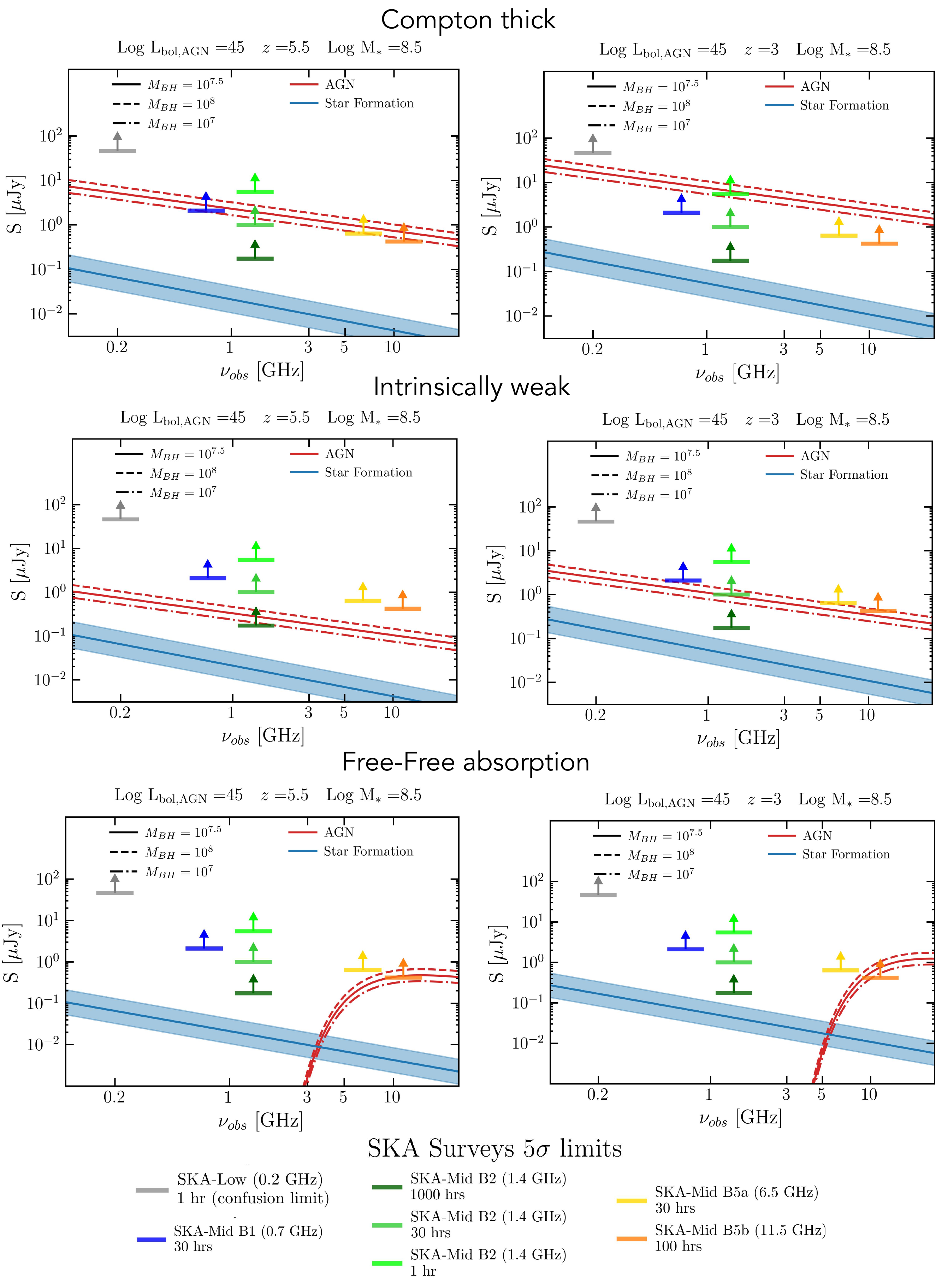}%{figures/panel_RQAGN.png}
    \vspace{-0.3cm}
    \caption{Predicted radio SED for the three main scenarios that can characterize LRD and JWST discovered AGN compared to the SKAO $5\sigma$ detection limits in the different bands. All quantities presented in the panels assume the AGN bolometric luminosity, redshift, and host stellar mass reported above each panel. Red lines represent the expected AGN radio fluxes computed assuming $M_{BH}=10^{7.5}$ \Msun, $\alpha=-0.5$ and for each scenario the procedure described in Sect.~\ref{sec:SKA}. The dashed and dash-dotted red lines refer to $M_{BH}=10^8$ \Msun and $M_{BH}=10^7$ \Msun, respectively. For the electron scattering scenario we assumed $T_e=5\times10^5~$K, $r=0.01~$pc and $N_e=10^{23}\rm ~ cm^{-2}$. The blue shaded area represents the range of radio fluxes associated to SF in the host galaxy (assuming the main sequence). The SKAO sensitivities are reported at 5$\sigma$ as a coloured thick with different colors for different bands and observing times. The gray thick represents the detection limit for 1 hour of observation with SKA-Low ($\sim0.2~$GHz), the blue thick the detection limit for 30 hours of observations in Band-1 ($\sim 0.7~\rm GHz$), the three green arrows represent the Band-2 (1.4~GHz) detection limits considering 1, 30, and 1000 hours of SKAO observations, the yellow thick the Band-5a (6.5~GHz) for 30 hours of observing time, and the orange one the Band-5b (11.5~GHz, ) detection limit for 100 hours of observations.}
    \label{fig:detectability}
\end{figure}

\subsubsection{Comparison with SKAO sensitivities}

As reported in Sect.~\ref{sec:intro}, recent works are also exploring the identification of analogs of these JWST-AGN and LRD also at $z<3$ and even in the local Universe \citep{Juodzbalis24_rosetta, Wang24, Loiacono25, Lin26_localLRD, Ma26_deep, Torralba26_LRDz17}. The LRD number density was initially found to decrease moving from the high-z to the lower-z Universe \citep{Ma26_deep}, while recent works revealed a flatter trend with redshift \citep{Loiacono25} and a larger low-z LRD population. In any case, these works unequivocally found that these sources (with the same physical, multiband, and spectroscopic properties) also populate a much younger Universe. Therefore, we decided to simulate the same three scenarios at two different redshifts, $z=5.5$ (the median redshift of most of the sample of JWST-discovered AGN, so far), and redshift $z=3$. 
%in the right column of Fig.~\ref{fig:detectability}.

%In Fig.~\ref{fig:detectability} we show the simulated AGN and host galaxy spectra, compared to the 5$\sigma$ detection limits that SKAO is expected to achieve at different observing times. {\bf All the sensitivities were computed using the online sensitivity calculator\footnote{\url{https://sensitivity-calculator.skao.int/}}, assuming the AA4 SKAO array configuration. In particular we assumed \texttt{Briggs} weighting and \texttt{robust}$=0$ and took the default full bandwidth for the continuum sensitivity.}

We also simulated the host galaxy radio emission that can contribute to the detectability of these sources, but also potentially overwhelm the AGN-related radio emission. We made the assumption of a stellar mass of the host $M_{*}=10^{8.5}$, representing the average of the range of stellar masses derived for these objects in the literature \citep{Greene24_LRD,Juodzbalis25_AGNsample,Mazzolari26_NOEMA, Ma26_undermasivehost}.  We derived the corresponding SFR using the main sequence from \cite{Popesso23}:
\begin{equation}
\log_{10} \mathrm{SFR} =
a_0 + a_1 t -
\log_{10} \left[
1 + \left(
\frac{M_*}{10^{a_2 + a_3 t}}
\right)^{-a_4}
\right],
\end{equation}
and then the radio flux associated with it using \cite{novak17} and the $q_{TIR}$ expression derived in \cite{delvecchio21}:
\begin{equation}
\begin{aligned}
L_{\mathrm{1.4~GHz}} =
\frac{\mathrm{SFR} \cdot 10^{24}}{f_{IMF} \cdot 10^{q_{\mathrm{IR}}}}
\, \left[\mathrm{W\,Hz^{-1}}\right], \quad
q_{\mathrm{IR}}(M_*, z) &= 2.646\,(1+z)^{-0.023}
- 0.148\,\left[\log_{10}(M_*) - 10\right].
\end{aligned}
\label{eq:novak_qtir}
\end{equation}
We then derived the radio flux associated with star formation using Eq.~\ref{eq:Lrad} and assuming a radio spectral index of $\alpha_{SF}=-0.7$, typical of star formation synchrotron emission. We also translated the main sequence relation scatter ($\sim 0.3~$dex) into a scatter in radio flux.

The results of the three scenarios presented above, compared with the $5\sigma$ sensitivities reached by SKAO at the different SKA-Low SKA-Mid frequencies and with surveys of different depths, are presented in Fig.~\ref{fig:detectability}. We approximately took as a reference the exposure times/sensitivities reported in \cite{Prandoni01.2026.SKA} for the Wide, Deep, and Ultra-Deep surveys.\\
All the sensitivities were computed using the online sensitivity calculator\footnote{\url{https://sensitivity-calculator.skao.int/}}, assuming the AA4 SKAO array configuration, a Precipitable Water Vapor of 10 mm, and a source at 45$^\circ$ of elevation. In particular, we assumed \texttt{Briggs} weighting and \texttt{robust}$=0$ and took the default full bandwidth for the continuum sensitivity. We considered all the currently planned Bands of the SKA-Mid (Band-1, Band-2, Band5a, Band5b), and also the SKA-Low band (centered at 0.2~GHz). The sensitivities reached with 1 hour of observation in the different bands are reported in Table~\ref{tab:sensitivities}

\begin{table}[ht]
\centering
\caption{Summary of SKAO bands' sensitivities with 1 hour of observation and achievable angular resolution for a target at 45$^\circ$ of elevation in the GOODS-S field ($\rm RA\sim 53^\circ$, $\rm DEC\sim -27^\circ$). ($^{**}$) We note that the Band 5b will benefit from an additional 64 receivers, not included in the SKAO sensitivity calculator, and that are expected to increase the sensitivity by a factor $\sim1.5$ compared to what is reported here.}
\begin{tabular}{cccc}
\hline
Band & Central Frequency [GHz] & $\rm rms$ [$\mu$Jy] (1 hr) & angular resolution [$^{\prime \prime}$] \\
\hline
% Example rows (replace with your values)
SKA-Low & 0.2 & 9.3 & 8\\
SKA-Mid Band 1 & 0.8 & 2.3 & 1.15\\
SKA-Mid Band 2 & 1.4 & 1.1 & 0.7\\
SKA-Mid Band 5a & 6.5 & 0.7 & 0.12\\
SKA-Mid Band 5b$^{**}$ & 11.5 & 0.84 & 0.07\\
\hline
\end{tabular}
\label{tab:sensitivities}
\end{table}

As can be seen from Fig.~\ref{fig:detectability}, for the first scenario (standard AGN emission but large hydrogen column densities to suppress the X-rays) Band-2 ($\sim \rm 1.4~GHz$) and Band-5a ($\sim \rm 6.5~GHz$) continuum observations of $\sim 20-30$ hrs are expected to detect the radio emission of $z\sim5.5$ sources with $S/N>5$. Instead, to detect them at $\sim 11~\rm GHz$ (Band-5b) will require exposures of $\sim 100$ hrs, while at lower frequencies (Band-1, $\sim 0.7~\rm GHz$) observations longer than $30$ hrs will be needed. 
%On the contrary, the AGN radio emission in the “weak” scenario is expected to be diluted by the SF contribution of the more massive host galaxies, and in general, to be detected only with the longest SKA integrations.
At $z\sim3$, it will be enough to observe these sources with only a few hours in Band-2 or in Band-5a, or a few tens in Band-1, to detect the radio emission predicted by this scenario.

In the intrinsically weak scenario, longer observations will be needed to detect the expected radio emission of LRD and JWST-AGN. At $z=5.5$ only the deepest observations in Band-2 (i.e. $\sim 1000$ hrs, those expected only for the Ultra-Deep survey, see \cite{prandoni15}) will be able to unveil their emission. Instead, at $z=3$, they can be detected with observations of several tens of hrs ($>30$ hrs) in Band-2 or Band-5a.

Finally, in the third scenario, where the radio emission comes from a standard AGN but partially suppressed by free-free absorption, the frequencies $\nu_{obs}<3-4\rm~ GHz$ are expected to be significantly impacted and the detectability will be possible only in the highest frequency bands, requiring $\gtrsim$30 hrs in Band-5a and $\gtrsim$100 hrs in Band-5b (both at $z\sim 5.5$ and at $z\sim 3$). We note that the detectability of radio emission from AGN surrounded by this ionized medium is highly sensitive to the free electron column density. If a column density of $N_e=10^{24}$ cm$^{-2}$ is assumed (keeping the other parameters fixed as in Sect.~\ref{sec:radio_detection}), the detectability of these sources becomes almost impossible due to the dependence of $\tau_{ff}$ on the second power of $N_e$.

Unfortunately, with the assumptions reported above, none of the three scenarios will be detectable at the sensitivities of the SKA-Low band ($f_{lim}=9.3\mu$Jy), where the confusion limit will be reached with just $\sim1$ hour of observation due to the large angular resolution ($\sim 8^{\prime\prime}$).

As shown in Fig.~\ref{fig:detectability}, the three scenarios predict very different observed radio fluxes and radio shapes. Therefore, thanks to the combination of SKAO observations across different bands and to the use of the predictions reported here, it will be possible to disentangle the different scenarios based on what deep SKAO continuum observations will detect (or not).

It is worth noting that in Sect.~\ref{sec:radio_detection} and in Fig.~\ref{fig:detectability} we considered the expected radio fluxes for single sources. However, by the time the SKAO becomes fully operational, several hundred (if not thousands) of spectroscopically confirmed AGN and LRD will be available (and much more photometric candidates), and stacking experiments will allow us to reach sensitivities $>10\times$ fainter than those reported in Fig.~\ref{fig:detectability} and much lower than the current attempts (see Tab.~\ref{tab:samples}).

Additionally, in all three scenarios, the AGN radio emission is expected to largely dominate the SF component, if the host is a main-sequence galaxy with $M_{*}\sim 10^ {8.5}$\Msun. The only exception is the frequency range where free-free absorption significantly affects the AGN emission in the third scenario.

It is worth noting that the predictions on the radio flux reported in Fig.~\ref{fig:detectability} assume a radio AGN spectral index of $\alpha=-0.5$. The possibility of detecting radio sources at different frequencies allows the radio spectral index to be constrained, which in turn provides important information on the origin of the emission. A flatter spectral index is indicative of radio emission from a compact core, as is typically observed in AGN \citep{Sina25_RQcompact}. This flat spectrum is thought to result from the superposition of multiple self-absorbed components, for example, within the AGN jet or the core of the AGN radio emission close to the central BH \citep[see][for a review]{panessa19}. In radio-quiet quasars, this can manifest as compact, flat-spectrum cores due to synchrotron self-absorption \citep{Laor08}. For example, the core of the high-redshift blazar GB 1508+5714 has a flat radio spectral index of 0.02, while its extended jet features have a steeper index of around -1.2 \citep{Kappes22}. Flatter radio spectral indices make it easier to detect sources at higher frequencies, as the flux density decreases more slowly with increasing frequency. On the other hand, a steep spectral index ($\alpha < -0.5$) is often associated with older, relic plasma in the extended structures of radio galaxies, or in general with an aged relativistic electron population (since the most energetic electrons, associated with the higher frequencies, are expected to lose their energies earlier). While steep radio spectra would favor low-frequency detections, steep-spectrum radio sources are not expected to be dominant in the high-z Universe due to the young BH ages.

%However, the possibility of detecting these sources at the different frequencies will also allow us to constrain the radio spectral index, deriving important information on the mechanism at its origin. Indeed, a flatter spectral index would be indicative of radio emission coming from a dense, partially absorbed emitting region, as typically observed in the cores of jetted and non-jetted AGN. Furthermore, flatter radio spectra indices would make the detection at higher frequencies easier. However, given the high redshifts of these sources, their emission might be young and therefore characterized by steeper spectral indices, making detection at higher frequencies harder.

The SKAO AA4 is expected to achieve sub-arcsecond resolution, reaching $\sim 0.7^{\prime\prime}$ at $\sim$1 GHz (Band-2) and $\sim 0.05^{\prime\prime}$ at $\sim$11 GHz (Band-5b)\footnote{Synthesized beam sized obtained assuming a source in the GOODS-S field with an elevation of 45$^\circ$}. Therefore, it will be possible to approximately match or even achieve better spatial resolutions than JWST/NIRCam  ($0.03^{\prime\prime}<\rm PSF_{NIRCam}<0.15^{\prime\prime}$, ranging between F070W and F444W filters\footnote{\url{https://jwst-docs.stsci.edu/jwst-near-infrared-camera/nircam-performance/}}) and NIRSPec-IFU observations ($0.16^{\prime\prime}<\rm PSF_{NIRSpec-IFU}<0.22^{\prime\prime}$, rainging between $1\mu m<\lambda<5\mu m$, see \citealt{Jones26_LRDz8}). This will allow us to test whether the radio emission of these sources is unresolved or not, and whether it matches the compact-sized emission observed in JWST images. In particular, it will be possible to test if there are any diffuse components coming from a faint host galaxy (too faint to be detected in the optical bands) or if there are any signatures of present or past radio jets expanding beyond the sub-kpc scale of the sources.

In case of non-detection of these sources, even with the longest SKAO observations, it will be possible to perform radio stacking analyses following the same attempts already performed in the literature (see Tab.~\ref{tab:samples}), but reaching radio flux densities two or three orders of magnitude fainter, thanks to both the deeper images and the larger number of available sources at that time. These experiments will also take advantage of the wide area covered by SKAO observations and by the larger number of JWST-selected AGN that can fall on these images.

\subsection{Variability}

The possibility of performing different tiers of observations at a constant periodicity will also enable us to test the variability of these sources in the radio band once they are detected. So far, studies analyzing NIRCam multi-epoch images for continuum variability in LRDs have found only weak or no variability \citep{Kokubo24, Tee24, Stone25,Zhang25_LRDvariab}. Given the peculiar physical properties of LRDs, detecting variability in multi-epoch observations would be an unambiguous sign that they are indeed AGN, because AGN continuum and broad lines are well known to vary over a broad range of observable timescales from hours to decades, while stellar populations do not. So far, the only LRDs (with multiple spectroscopic follow-up) for which spectral variability was detected are QSO1a at $z\sim 7.04$ \citep{Ji25, Furtak25} and the $z\sim7$ LRD reported in \cite{Lambrides26_LRDvariab}.

\begin{figure}[h!]
    \centering
    \includegraphics[width=0.9\linewidth]{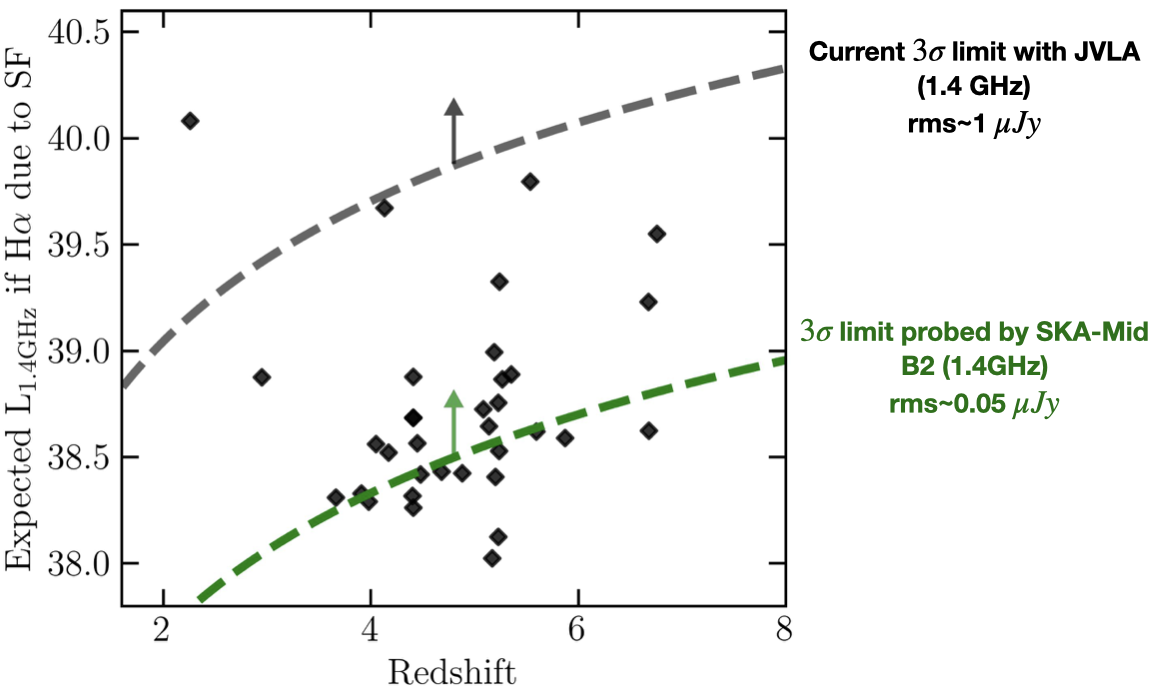}
    \caption{Expected 1.4~GHz radio luminosity versus redshift of the JWST discovered Type I AGN on GOODS-N field, assuming the whole \Ha emission (broad + narrow) to be due to star formation. The radio luminosity values are compared to the GOODS-N VLA detection limit ($\sim 1 \mu $Jy) and the SKAO Band-2 detection limits ($\sim$20 times fainter). It will be possible to test the non-AGN scenario for the bulk of the sources.}
    \label{fig:SF_radio}
\end{figure}

The lack of observed rest-frame UV variability has been explained by high AGN luminosity or strong galaxy contamination \citep{Zhou25}, or by ongoing super-Eddington accretion, which is expected to produce a lower amount of variability due to the fact that UV emission from the AGN is absorbed or scattered \citep{Secunda25}. \\
Different works highlighted similarities between the JWST-discovered AGN and the well-known, already-explored population of Narrow-line Seyfert I galaxies \citep{Mathur00, Zhou06}. This is a population of highly accreting AGN characterized by BH masses similar to those observed in high-$z$ JWST AGN (i.e. $10^6<M_{BH}<10^8$), that often accrete close or above the Eddington ratio \citep{Tortosa22,Tortosa23b}, and by a steep and soft X-ray spectrum \citep[determining an apparent X-ray weakness][]{Gallo18}. This population shows a variety of different behaviors in the radio band, ranging from strong radio jets and flares \citep{Lahteenmaki18,Jarvela21}  to the (dominant) radio-quiet behavior \citep{Berton16, Berton18, Rakshit17,Singh18,Varglund25}, and the fraction of NLS1s detectable in radio greatly depends on the sample, the frequency, and the survey \citep{Berton25} and ranges from a few percent up to $\sim 30\%$. Some of these sources are also characterized by fast radio variability \citep{Jarvela24}, which is thought to be associated with the interaction between small-scale jets and the BLR clouds \citep{Delpalacio19}, or with magnetic reconnection phenomena in the black hole magnetosphere \citep{Kadowaki15}. Given the similarities between NLS1 and JWST-discovered AGN, it is possible that the latter class of objects can also undergo similar phenomena and therefore show a wide variety of behaviors in the radio band, including possibly significant time variability.

The only LRD analog detected so far in the radio band (and residing in the local Universe, $z=0.168$) was observed at similar radio frequencies in 2010 and 2018, without revealing significant variability in the radio flux. \cite{Rodriguez26_radioLRD} argued that the lack of variability in the radio band can exclude the radio emission from coming from a single supernovae event (which would have implied a variation in the radio flux within the two observing dates), while it instead favors the BH origin.

\section{Test Alternative scenarios}\label{sec:alternative}
%\subsection{High stellar densities and tidal disruption events}
It has also been proposed that the broad Balmer lines characterizing JWST-discovered AGN and LRDs, instead of being attributed to AGN BLR emission, are due to tidal disruption events \citep{Bellovary25} or supernovae, i.e., transient events that can produce similarly broad emission components in the Balmer lines.
The tidal disruption events (TDE) scenario assumes that the broad Balmer lines arise from the disruption of stars that live in runaway-collapsing star clusters. \cite{Bellovary25} suggested that, to match the LRD number density, a rate of 10$^{-4}$ TDE events per year is necessary.
Recently, \cite{Perger25} tested the TDE scenario in light of the radio non-detection of these sources. They considered two different scenarios for the radio emission of TDEs and concluded that the non-detection of these objects in the radio band is compatible with RQ TDE phenomena.
However, large uncertainties remain for this scenario. Indeed, while TDEs might be able to explain also the X-ray weakness of these sources, they are not able to explain most of the other LRD properties, such as the red colors, the V-shape, or the strong Balmer break.\\

The possibility that single or cumulative supernova events can justify the properties of these sources was discussed in different works \cite{Matthee23,Zhang25_LRDvariab,Maiolino24_X}. In particular, \cite{Maiolino24_X} showed that the supernova scenario is unlikely because the observed broad \Ha luminosities are significantly higher than the cumulative light expected from multiple supernovae at the inferred star formation rates of the host galaxies. Furthermore, the general lack of variability in multi-epoch observations \citep{Zhang25_LRDvariab} is inconsistent with individual supernova events, while other spectral properties typical of SNe, such as P-Cygni profiles or additional broad metal lines, are not observed in the JWST spectra.

\cite{Baggen24} suggested that the broad Balmer lines could also be explained in a purely stellar scenario. Due to the small sizes of these sources, the derived stellar masses would imply high stellar densities (similar to those of the densest nuclear star clusters) and rapid stellar motion, which could explain the broad lines without the need to invoke emission coming from the rapidly rotating clouds of the broad line region around an active BH. However, if all the LRD light is dominated by star formation, LRDs on their own can account for (or even exceed) the predicted stellar mass functions for all galaxies at high redshift in $\Lambda$CDM \citep{Boylan-Kolchin23,Akins25_LRD}, unless nonstandard, top-heavy initial mass functions (IMFs) are assumed.

Deep radio observations can further test whether the nuclear star cluster scenario, also assumed by the TDE hypothesis, or the supernovae scenarios are realistic. Assuming that both the broad and narrow H$\alpha$ emission in these objects comes only from star formation (instead of from the AGN BLR and NLR), it is possible to derive the corresponding SFR using the relation reported in \cite{Shapley23, Reddy22}: $\mathrm{SFR} = 10^{-41.67} \, L_{\mathrm{H}\alpha}
\quad \left[M_\odot\,\mathrm{yr}^{-1}\right],$ and then translate it into a 1.4~GHz radio luminosity using Eq.~\ref{eq:novak_qtir}. We considered the sample of Type I AGN discovered by JWST in the GOODS-N field, and we computed the expected radio luminosity under the assumption of pure SF. The results are reported in Fig.~\ref{fig:SF_radio}. Sources above the detection thresholds should be those detectable in the radio band if the pure stellar scenario is valid. None of these sources is radio-detected in the deep GOODS-N images, but as it is possible to see in Fig.~\ref{fig:SF_radio}, with current radio surveys, we can rule out the nuclear star cluster scenario only for a couple of sources. Instead, with the deepest SKAO observations, it will be possible to potentially rule out (or support) this scenario for the bulk of these sources (probing SFR of the order of  $5-50\, \Msunyr$~ at $z\sim 5-6$).

\section{Survey Strategy and Conclusions} \label{sec:conclusion}
To pursue the goals proposed in Section~\ref{sec:SKA} and Sect.~\ref{sec:alternative}, we need deep radio observations in the extragalactic fields observed with JWST and accessible to the SKAO, such as GOODS-S and COSMOS, and the lensed fields in the Southern Hemisphere. In Sect.~\ref{sec:SKA} we reported the expected AGN radio emission versus radio frequencies spanned by the SKA-Low and Mid observations, assuming three different scenarios that can explain the overall emission of JWST discovered AGN and LRD: a standard AGN with a Compton-thick envelope, an intrinsically weak (in X-ray and radio) AGN, or a standard AGN with an ionized envelope and subject to free-free absorption.  The three scenarios are responsible for very different emissions across frequencies, and therefore, deep, multi-frequency SKAO observations will be able to shed light on the real physical nature of these sources. In particular, to detect AGN described by the first scenario, it will be enough to have observations of a few hours in SKA-Mid Band-2 ($\sim$1.4~GHz) at $z\sim3$, and a few tens of hours for sources at $z\sim5$. A few tens of hours will also be enough to detect them in SKA-Mid Band-5a ($\sim8$ GHz).\\
In the intrinsic week scenario, the deepest observations will be needed to detect the AGN radio emission. At $z\sim3$, $T_{obs}>30$ hours in Band-2 and Band-5a and $T_{obs}>100$ hours in Band-5b will be needed, while at $z\sim5$, a detection will only be possible with $\sim 1000$ hours of Band-2 observations. \\
In the ionized cocoon scenario, a detection at frequencies $<3$GHz will be prevented by free-free absorption, while it will be possible to detect these AGN with $>30$ and $\sim 100$ hours of observations in B5a B5b (not affected by free-free absorption).\\

The possibility of covering the same fields with radio observations at different SKAO frequencies will enable us to perform radio spectral index analyses, deriving crucial information on the physical mechanisms powering the radio emission of these sources, as described in Section~\ref{sec:SKA}.

To explore the potential radio variability of these sources, we also need observing tiers with a periodic cadence, for example, of 2–3 months, following the example of the NEXUS JWST survey \citep{Stone25, Zhuang25}.\\

Finally, thanks to the depth of the SKAO continuum survey it will be also possible to test alternative scenarios to BH accretion as the origin of the intrisic power of these sources. In particular, it will be possible to test whether these systems are powered by nuclear star cluster or supernovae.

%We expect these observations to be organized in tiers, with well-defined time intervals between observations, and possibly considering at least two different radio frequencies available with SKA-Mid. Considering Fig.~\ref{fig:detectability}, to detect the radio emission from these sources we might need the deepest radio observations that SKAO will perform, i.e. depths similar to those of the Ultra-deep survey according to \cite{prandoni15}. To explore the potential radio variability of these sources, we also need observing tiers with a periodic cadence, for example, of 2–3 months, following the example of the NEXUS JWST survey \citep{Stone25, Zhuang25}. Lastly, the possibility of covering the same fields with radio observations at different SKAO frequencies will enable us to perform radio spectral index analyses, deriving crucial information on the physical mechanisms powering the radio emission of these sources, as described in Section~\ref{sec:SKA}. Additionally, the possibility to observe at the highest frequencies of the SKA-Mid would possibly prevent the undetection of these sources as a consequence of free-free absorption of the radio emission, which may arise in these objects as a consequence of the presence of large resevoirs of ionized hydrogen atoms and free electrons.

\section*{Acknowledgements}
GM and HÜ acknowledge funding by the European Union (ERC APEX, 101164796).

\bibliographystyle{abbrvnat-maxbibnames4}
\bibliography{chapter} % if your bibtex file is called example.bib

\end{document}